\documentclass[proof]{WileyASNA-v2}

\articletype{Article Type}%

\received{xxx}
\revised{xxx}
\accepted{xxx}

\begin{document}

\title{Resolving stellar populations with integral field spectroscopy}

\author[1,2]{Martin M. Roth*}

\author[1]{Peter M. Weilbacher}

\author[1]{Norberto Castro}

\authormark{Roth \textsc{et al}}

\address[1]{\orgname{Leibniz-Institut f\"ur Astrophysik Potsdam (AIP)}, \orgaddress{\state{Potsdam}, \country{Germany}}}

\address[2]{\orgdiv{Institut für Physik und Astronomie}, \orgname{Universität Potsdam}, \orgaddress{\state{Potsdam}, \country{Germany}}}

\corres{* \email{mmroth@aip.de}}
\presentaddress{Leibniz-Institut für Astrophysik Potsdam (AIP), An der Sternwarte 16, 14482 Potsdam, Germany}

\abstract{High-performance instruments at large ground-based telescopes have made integral field spectroscopy (IFS) a powerful tool for the study of extended objects such as galaxies, nebulae, or even larger survey fields on the sky. Here we discuss the capabilities of IFS for the study of resolved stellar populations, using the new method of PSF-fitting crowded field IFS, analogous to the well-established technique of crowded field photometry with image sensors. We review early pioneering work with first generation integral field spectrographs, the breakthrough achieved with the MUSE instrument at the ESO Very Large Telescope, the remarkable progress accomplished with MUSE in the study of globular clusters, and first results on nearby galaxies. We discuss the synergy of integral field spectrographs at 8-10m class telescopes with future facilities such as the Extremely Large Telescope (ELT).
}



\maketitle

\section{Introduction}\label{sec1}

Integral field spectroscopy (IFS) is a technique that was developed in the early 1990s to address the need of measuring spectra for spatially extended objects, in particular galaxies, and to overcome the limited coverage of long slit spectrographs, let alone the one of optical fiber coupled spectrographs. Scanning long slit spectroscopy or scanning Fabry Perot imaging were found to be too expensive in terms of observing time, and less than ideal due to variable atmospheric conditions over the course of an observation of a given target. Reviews of integral field spectroscopy and applications of this technique are found in \citet{Roth2010} and \citet{Eisenhauer2015}. An excellent overview is given in \citet{Bacon2017b}.

The main momentum for IFS was created from demands in extragalactic astronomy, where it was discovered that in order to understand the formation and evolution of galaxies the two-dimensional and spatially resolved coverage of entire galaxies would be essential. For example, despite the undisputed impact of the Sloan Digital Sky Survey and its instrumentation  \citep{York2000}, the fiber aperture of $\approx$ 3~arcsec diameter is too small to sample the extent of galaxies sufficiently well. First generation integral field units (IFU) of instruments such as e.g. SAURON \citep{Bacon2001} or PMAS \citep{Kelz2006,Roth2005} have enabled studies with statistically meaningful samples like the SAURON survey \citep{deZeeuw2002}, ATLAS3D \citep{Cappellari2011}, and CALIFA \citep{Sanchez2012}, that have shed light on the evolution of elliptical and disk galaxies, respectively, including e.g. the discovery of different classes of fast/slow rotators \citep{Emsellem2011}, arguments on the inside-out scenario for star formation histories, local mass-metallicity-relations, etc. \citep{Sanchez2019}.

The advent of MUSE as a second generation instrument for the ESO Very Large Telescope (VLT) has further boosted the capabilities. The IFU features a 1$\times$1 arcmin$^2$ FoV with spatial sampling of 0.2~arcsec, high throughput, excellent image quality, and a one octave free spectral range with a spectral re\-so\-lution
of R = 1800 \ldots 3600 \citep{Bacon2014}, that have enabled, amongst others, remarkable discoveries in the high redshift universe
\citep{Bacon2017,Wisotzki2018}.

In this paper, we review the unique opportunities offered by crowded field integral field spectroscopy with instruments such as MUSE to study resolved stellar populations in the Milky Way and nearby galaxies.

\section{The technique of PSF-fitting crowded field spectroscopy}\label{sec2}

\begin{figure}[t]
	\centerline{\includegraphics[width=88mm]{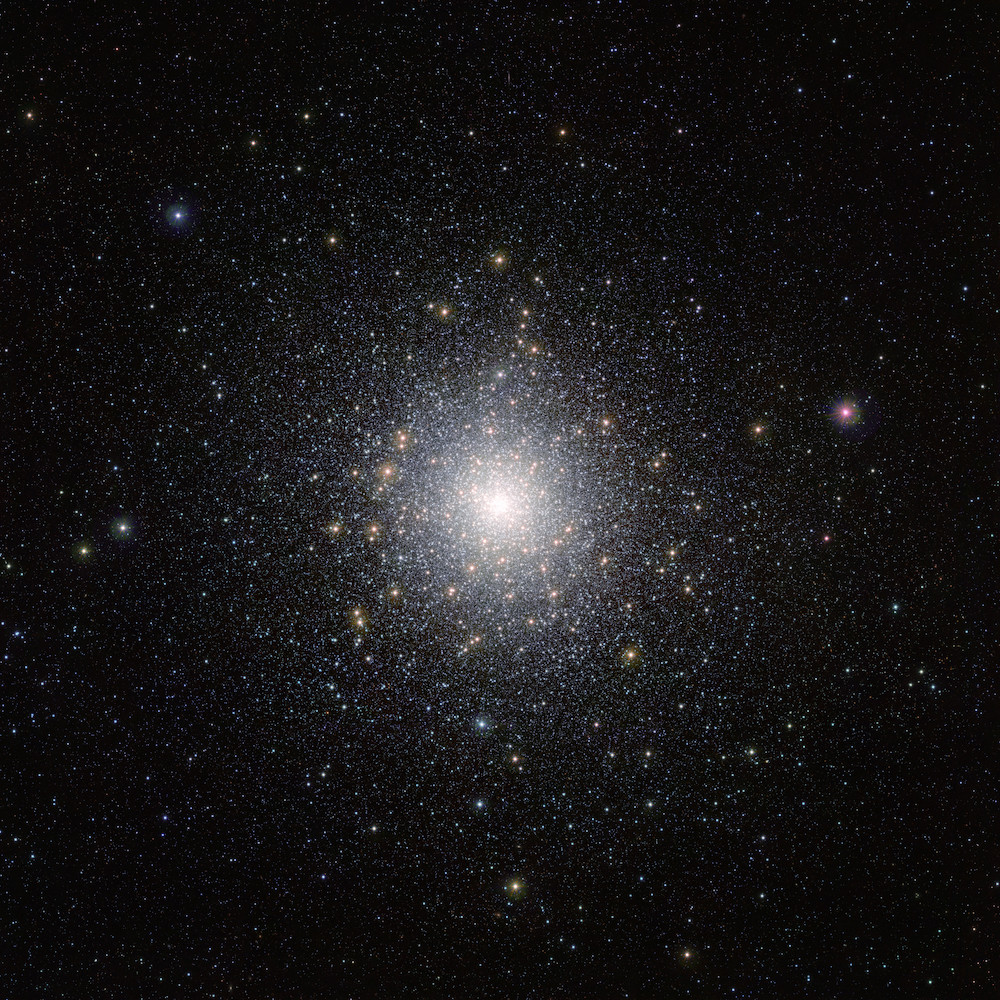}}
	\caption{The globular cluster 47 Tuc. Credit: ESO/M.-R. Cioni/VISTA Magellanic Cloud survey. \label{47Tuc}}
\end{figure}

Crowded stellar fields, e.g. in the dense cores of globular clusters (GC), have remained inaccessible to photometry before the advent of image sensors, e.g. the CCD detector, allowing for a digital recording with a linear response and a large dynamic range, in combination with image processing software, e.g. DAOPHOT \citep{Stetson1987}. A an example, Figure~{\ref{47Tuc}} illustrates the severe blending of stellar images in the nucleus of 47~Tuc to the point where individual stars can no longer be distinguished. However, DAOPHOT (and other computer programs), utilizing knowledge of the point-spread-function (PSF) for a given image, have demonstrated that stellar images can be deblended through simultaneously fitting the PSF to the centroid of each star, even when these images are significantly overlapping with each other. The tremendous progress enabled by the transition from photographic/photoelectric photometry to crowded field CCD photometry can hardly be exaggerated, and is probably best  appreciated for the example of 47~Tuc by comparing the respective colour-magnitude diagrams presented in \citet{Hesser1987}, their Fig.~9.

Stimulated by early work with IFS in crowded fields as described in Section~\ref{sec3}, \citet{Kamann2013a}\footnote{available online: https://d-nb.info/1042331073/34} undertook the task to develop a PSF fitting code for datacubes, analogous to DAOPHOT for two-dimensional images, for the purpose of studying the nuclei of GC and the presence of putative intermediate mass black holes. The program is called PampelMuse (Potsdam Advanced Multi-PSF Extraction Algorithm for MUSE) and is publicly available
\citep{Kamann2018c}
\footnote{https://gitlab.gwdg.de/skamann/pampelmuse}.

PampelMuse works on the assumption that the centroids of all stars in the field with appreciable flux are known a priori, to be provided as an input source catalogue.  In practical terms, this is best accomplished with existing archival Hubble Space Telescope (HST) images that have a sufficiently high angular resolution and large dynamic range. The PSF is modeled with a Moffat function, that accounts for non-circular images and extended wings, compared to a simple Gaussian fit. The wavelength-dependence of the PSF is determined through all layers of the datacube inside-out, and applying the constraint that the PSF parameters must be smoothly varying functions of wavelength. 

The algorithm compares an observed datacube:
\begin{eqnarray}
D_{i,j,k}
\end{eqnarray}
noting subscripts $i,j$ for the spatial coordinates, and $k$ for wavelength, with a global model datacube $M$, written as the sum of stars $n$ with flux contributions $f$ across the point-spread-function $P$, and a background term $B$ with subscript $m$ that is taking into account unresolved faint stars:
\begin{eqnarray}
M_{i,j,k} &= & \sum\limits_{n}   f_{k}^{n}  P_{i,j,k}^n + \sum\limits_{m} B_{i,j,k}^m
\end{eqnarray}
A best fit is then determined by minimizing:
\begin{eqnarray}
\chi^2\ &= &   \sum\limits_{i,j,k}   \frac{(D_{i,j,k} - \sum\limits_{n}   f_{k}^{n} P_{i,j,k}^n - \sum\limits_{m} B_{i,j,k}^m)^2}{\sigma_{i,j,k}^2}
\end{eqnarray}
The code was tested extensively with simulated datacubes that were created for 47~Tuc from published HST images and photometry, a single isochrone of given age and metallicity, and spectra selected from a library for 
T$_{\mathrm{eff}}$ and log~$g$. The spectra were also Doppler-shifted for a random distribution of radial velocities.
A full account of details of the algorithm, as well as test results from the simulations, decribing the performance with regard to cross-talk, radial velocities, and equivalent widths, the effect of signal-to-noise ratios (SNR) and of crowding, etc. is given in \citet{Kamann2013b}.

As the acronym implies, PampelMuse was intended to become a tool for the MUSE instrument \citep{Bacon2014}, that was still under construction by the time the code was written. Therefore, an on-sky validation was first undertaken with PMAS at at the Calar Alto 3.5m telescope \citep{Roth2005}. Since then,  as reviewed in Section~\ref{sec4}, PampelMuse has been applied successfully to numerous studies of various kinds.


\pagebreak

\section{Early work with first generation integral field spectrographs}\label{sec3}
The capability of IFS for crowded field spectroscopy was first demonstrated using the first generation integral field spectrograph 3D \citep{Weitzel1996}. K band observations around the Galactic Center near Sgr A$^*$ revealed the presence of a cluster of massive, helium-rich blue supergiant/Wolf-Rayet stars and enabled to measure the kinematics of individual stars \citep{Krabbe1995}, thus paving the way to monitoring the orbit of objects like the star S2, and eventually measuring gravitational redshift with GRAVITY more than two decades later, i.e. providing direct evidence of the presence of a supermassive black hole in the Milky Way, and a demonstration of predictions from general relativity \citep{Gravity2018}. Although the images of the Galactic Center star cluster presented in the paper by \citet{Krabbe1995}  were actually not reconstructed from data of the 3D instrument, but rather obtained separately with the MPE speckle imaging camera SHARP \citep{Krabbe1993}, the plotted spectra provided ample evidence that IFS is indeed a powerful tool to resolve crowded stellar fields, despite the fact that PSF fitting techniques for datacubes where as yet not developed at the time.

However, the PMAS science case \citep{Roth1997,Roth1998} was strongly motivated by this idea, leading to experiments during the development phase of the instrument by using MPFS at the Selentchuk 6m telescope \citep{Silchenko2000}, that resulted in first encouraging results reported by \citet{Roth2000}.

The first paper published from PMAS data after commissioning has indeed demonstrated that PSF-fitting techniques known from direct imaging are capable to disentangle blended point sources, in this case for a quadruple gravitational lens \citep{Wisotzki2003}, followed by \citet{Christensen2003}, who reported integral field spectroscopy of SN2002er.

The capability of IFUs of providing accurate spectrophotometry by means of fitting the PSF to supernovae as point sources that are superimposed on the surface brightness of their host galaxy, was extensively exploited by the Nearby Supernova Factory collaboration, using SNIFS at the University of Hawaii 88-inch telescope on Mauna Kea
\citep{Childress2013,Copin2009}.

With data obtained from PMAS and MPFS for objects in the nearby galaxies M31 and M33, \citet{Becker2004} reported results from experiments with crowded field 3D spectroscopy, that employed, on the one hand, the image restoration algorithm {\it cplucy} \citep{Hook1994}, and, on the other hand, a Moffat fit to point sources through the layers of the respective datacubes.

Building on the encouring results from the latter approach, PMAS and MPFS observations of planetary nebulae in the bulge of M31 were analyzed with the PSF-fitting technique to resolve a problem with source confusion from the 
surface brightness background in M31 that was known from the literature \citep{Roth2004}. A similar approach was used successfully for the observation of LBV candidates in M33 \citep{Fabrika2005}.

In essence, these early attempts to exploit the two-dimensional information provided by IFS for crowded stellar fields have demonstrated two benefits: firstly, the multiplex obtained by covering an area rather than a slit (or the tiny aperture of a fiber), and, secondly, the suitability of the PSF-fitting technique to disentangle point sources from their environment. The small field-of-view of first generation integral field spectrographs, however, has prevented the technique from becoming a break-through at a larger scale. This situation has changed dramatically with the advent of MUSE at the VLT.

\begin{figure*}[th!]
\centerline{\includegraphics[width=17.8cm]{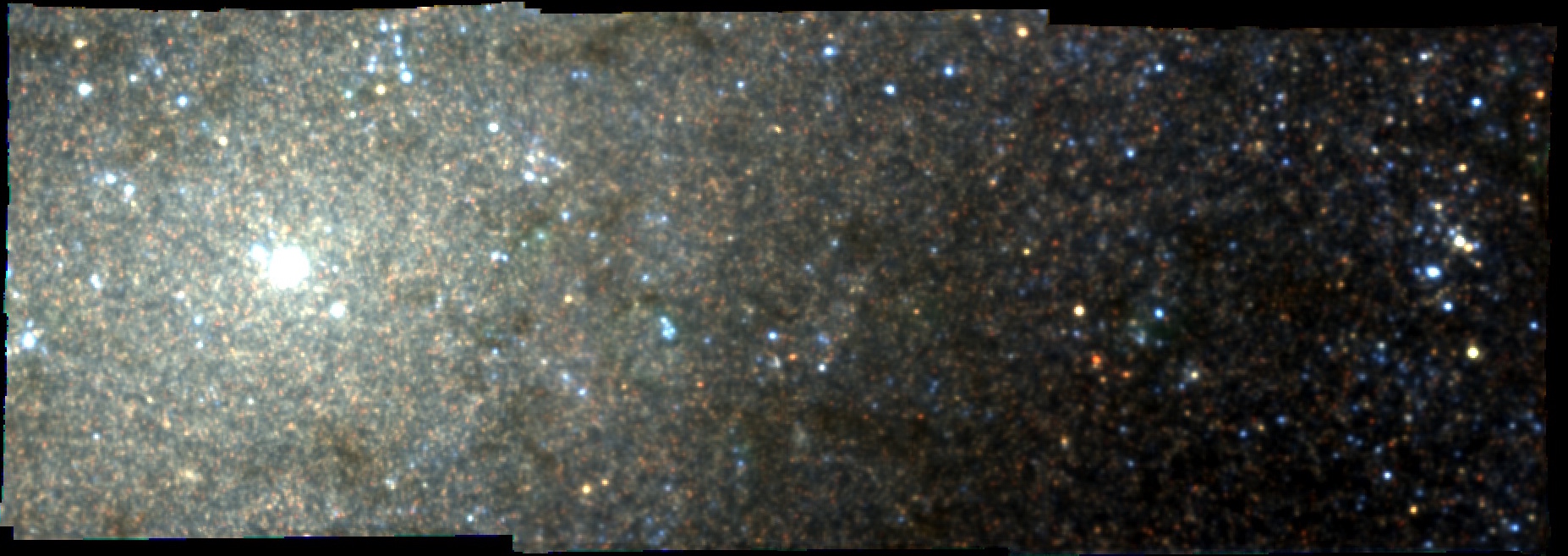}}
\centerline{\includegraphics[width=17.8cm]{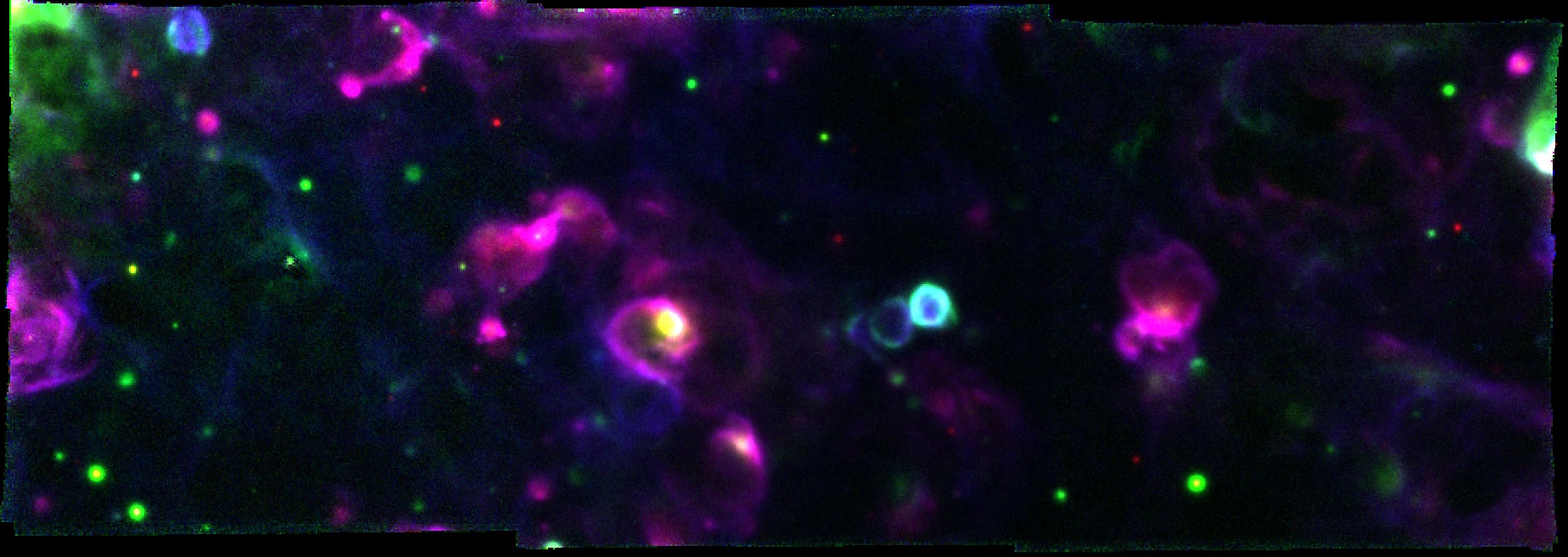}}
\caption{Stars and gas in NGC300, observed with three adjacent MUSE pointings from the nucleus of the galaxy to the West over a 3$\times$1~arcmin$^2$ field-of-view.
Image quality: 0.6~arcsec FWHM.
Top: RGB image created by collapsing the datacubes within V, R, I broadband filter curves. Bottom: composite narrowband images, highlighting H$\alpha$ in red, [OIII] in green, and [SII] in blue hues. Numerous distinct objects are immediately apparent, e.g. blue and red supergiants in the top panel, HII regions (purple), supernova remnants (blue),  planetary nebulae as green dots, and emission line stars as faint red dots in the bottom panel.
\label{NGC300}}
\end{figure*}

\section{Globular Clusters}\label{sec4}

In anticipation of MUSE, the PampelMuse tool was developed and tested in a pilot study using PMAS at the Calar Alto 3.5m telescope. Despite the smaller telescope aperture and comparatively small field-of-view, \citet{Kamann2014}
succeeded to mosaic the GCs M3, M13, and M92 around their nuclei, deblend a total of 225 stars using PampelMuse, and measure radial velocities for 564 spectra with uncertainties down do better than 1~km~s$^{-1}$, to obtain stringent limits on the masses of intermediate-mass black holes in those GCs.

A breakthrough, however, was accomplished upon the commissioning of MUSE at UT4 of the VLT observatory. A snapshot series of typically 1~min exposures, covering a mosaic of 5$\times$5 pointings centered on NGC~6397, yielded a total of 18932 spectra for 12307 stars, 10521 of which exhibit a SNR~$>10$.  From this data, \citet{Husser2016} derived a mean radial velocity of $v_{rad} = 17.84\pm0.07$~km~s$^{-1}$, and a mean metallicity of [Fe/H] = $-2.120 \pm0.002$ that seems to vary with temperature for stars on the red giant branch (RGB). 
T$_{\mathrm{eff}}$ and [Fe/H] were derived from the spectra, while log~$g$  was obtained from HST photometry. For the first time, a comprehensive Hertzsprung-Russell diagram (HRD) for a GC was created on the basis of the analysis of several thousands of stellar spectra, ranging from the main sequence to the tip of the RGB.

From the same dataset of more than 18000 spectra, \citet{Kamann2016} derived the kinematics of individual stars, reaching a typical radial velocity accuracy of 1~km~s$^{-1}$. The velocity dispersion profile was found to show a mild central cusp, and there was slight evidence for a rotational component in the cluster. By comparing spherical Jeans models with the kinematical data, and assuming a constant mass-to-light ratio, the presence of an intermediate-mass black hole with a mass of 600~M$_\astrosun$ was shown to be plausible.
Fostered by this early success, a comprehensive GC observing program was implemented in the guaranteed observing time (GTO) plan of the MUSE Consortium. This plan, although conducted in visitor mode, is a coordinated and efficient team effort to achieve best use of GTO through an internal priority schedule that meets the different requirements of various subprograms in terms of observing conditions such as image quality, moon phase, etc. The GC subprogram was therefore allowed to be scheduled in staggered cadence in order to address time series measurements and a search for spectroscopic variability.
Results from this program have been published in a series of papers:

\begin{itemize}
\item A stellar census in globular clusters with MUSE: The contribution of rotation to cluster dynamics studied with 
200\,000 stars \citep{Kamann2018b}
\item A stellar census in globular clusters with MUSE: A spectral catalogue of emission-line sources\citep{Gottgens2019b}
\item A stellar census in globular clusters with MUSE: Multiple populations chemistry in NGC 2808 \citep{Latour2019}
\item A stellar census in globular clusters with MUSE: Binaries in NGC 3201 \citep{Giesers2019}
\item Discovery of an old nova remnant in the Galactic globular cluster M 22 \citep{Gottgens2019a}
\item A detached stellar-mass black hole candidate in the globular cluster NGC 3201 \citep{Giesers2018}
\end{itemize}

Furthermore, \citet{Kamann2018a} were able to demonstrate with observations of the intermediate-age star cluster NGC~419 in the Small Magellanic Cloud that even rotational broadening of stellar spectra is measurable, despite the modest spectral resolution of MUSE.

After commissioning of the MUSE Narrow Field Mode (NFM) and science verification \citep{Leibundgut2019}, it has become apparent that adaptive optics further enhances the capability of MUSE for crowded field spectroscopy to the point of achieving an angular resolution similar to HST. 


In the meantime, other groups have discovered the benefits of MUSE crowded field spectroscopy. Some recent papers include the work of  \citet{Ernandes2019}, who have studied the metallicity of the GC Terzan~9, that at a distance of 
7.7~kpc from the sun and an orbit of less than 1~kpc from the Galactic Center is located in the inner bulge of the Milky Way. \citet{Alfaro2019} have observed the central cluster M54 of the Sagittarius dwarf galaxy with 4$\times$4 MUSE pointings, resulting in a total of 55000 spectra, {\textasciitilde}6600 of which were analyzed with ULySS \citep{Koleva2009} to show the presence of three distinct stellar populations, whose ages and metallicities were explained with accretion and merger of GCs, and the recent encounter of the dwarf galaxy in the process of merging with the Milky Way. As for another class of objects, \citet{Zeidler2018} have performed MUSE observations of the young massive star cluster Westerlund~2 to classify a total of 72 spectra into spectral types O (4), B (7), and A$\ldots$G (6), while 55 stars remained unclassified, and to measure radial velocities for all of the 72 spectra. It is useful to recall that all of these studies were performed by relying on PampelMuse as the tool to extract spectra from the datacubes.


\section{Nearby galaxies}\label{sec5}

Given the success of crowded field IFS with GC, a more ambitious goal would be to reach out to resolve stellar populations in nearby galaxies, such as the Magellanic Clouds, which are roughly an order of magnitude more distant than the clusters studied so far within the Galaxy. \citep{Castro2018} have indeed undertaken the task to  map the core of the Tarantula Nebula with MUSE in order to study stars and the nebular around the young cluster R136, which presents the strongest star formation activity in the local group, and is thought to contain the most massive stars in the nearby Universe. With 4 pointings at different exposure times, a total of 2255 spectra were secured, yielding 270 radial velocities for massive stars. The emission line gas diagnostics allowed to construct an extinction map and to trace the
the interstellar medium with new insights in regions where stars are probably being formed. 

Even further out at a distance of 1.6 Mpc,  \citet{Evans2019} observed the Leo P dwarf galaxy. With one deep pointing of 6.7~hours exposure time, they obtained a total of 341 spectra, 32 of which were classified to be of spectral type OB,
5 as AGB star candidates, and 6 of spectral types K$\ldots$M, while 298 spectra remained unclassified. The sensitivity of MUSE to faint emission lines led to the discovery of three 100 pc-scale ring structures in H$\alpha$, that were contemplated to represent old supernova remnants rather than wind-blown bubbles around massive stars.

Again as part of the MUSE Consortium GTO plan, observations of the 1.88~Mpc distant SA(s)d galaxy NGC\,300 were undertaken under conditions of excellent image quality, yielding a total of 9 pointings with expose times of 1.5~hours per field. A subset of three adjacent pointings, that were obtained with ground-layer adaptive optics support, are highlighted in Figure~\ref{NGC300}\!.  A pilot study using this data by \citet{Roth2018} has demonstrated that MUSE enables spectral type classification and the measurement of radial velocities of individual giants$\ldots$supergiants even in crowded fields of nearby galaxies beyond the local group. For the presentation of first results, merely one out of the 9 fields was fully analyzed for individual stars, yielding a total of 517 objects for which the SNR was good enough to allow a spectral type classification and measuring radial velocities. In addition to the stars, a rich inventory of other objects was collected, such as 45 planetary nebulae, 61 H\,II regions, 51 compact H\,II region candidates, 38 supernova remnant candidates, 118 emission line stars, and 30 background galaxies. Following up to this pilot study, several papers presenting the analysis of the full data set are in progress.


\section{Conclusions and outlook}\label{sec5}

The (incomplete) list of publications presented in this review of stellar spectroscopy performed to date with MUSE illustrates that the vision of crowded field IFS has become a reality. In view of the the fact that the study of resolved stellar populations represents one (of nine) major science cases for the ELT \citep{Hook2009}, it is reassuring to note that current work with MUSE at the VLT is paving the way for future studies exploiting the light collecting power of a high angular resolution 39m telescope with adaptive optics, and the integral field spectrograph HARMONI  as a first light instrument  \citep{Gonzalez2018}. Also synergy with MICADO and ELT-MOS seems to be obvious. It is worthwhile to note that the BlueMUSE instrument, that was proposed recently \citep{Richard2019}, would further enhance the capabilities of studying resolved stellar populations near and far.

\section*{Acknowledgments}

MMR, PMW, and NC acknowledge financial support from BMBF Verbundforschung grants  05A17BAA and  05A17BA1.


\nocite{*}
\bibliography{STARS2019_Roth}%

\end{document}